# Charge fractionalization in biased bilayer graphene

J. C. Martinez[1a], M. B. A. Jalil[1,2], S. G. Tan[2,3]

[1]*Information Storage Materials Laboratory, Electrical and Computer Engineering Department, National University of Singapore, 4 Engineering Drive 3, Singapore 117576*

[2]*Computational Nanoelectronics and Nano-device Laboratory, National University of Singapore, 4 Engineering Drive 3, Singapore 117576.*

[3]*Data Storage Institute, DSI Building, 5 Engineering Drive 1, National University of Singapore, Singapore 117608*

We study charge fractionalization in bilayer graphene which is intimately related to its zero modes. In the unbiased case, the valley zero modes occur in pairs rendering it unsuitable for charge fractionalization. A bias plays the role of a bosonic field with nontrivial topology allowing for the exploration of Dirac-like dynamics at higher particle momenta. It also induces an odd number of zero modes, which, together with the conjugation symmetry between positive and negative energy states, are the requisite conditions for charge fractionalization. A self-conjugate, localized zero mode is constructed for a semi-infinite graphene sheet with zigzag edge; scenarios can occur readily where one sublattice component (pseudospin) dominates. While the other valley also has a similar zero mode, a layer asymmetry can be invoked to lift this degeneracy allowing for detection of distinguishable charge-½ edge states per valley.

---

[a] Email: elemjc@nus.edu.sg

Fermion number fractionalization is a counterintuitive concept, first introduced by Jackiw and Rebbi [1] in the context of a fermion field coupled to a bosonic field with a nontrivial soliton profile, which has been extensively studied in field theoretic systems [2]. Although fractional excitations in polymers have been regarded as manifestations of fractionalization [3], the direct detection of fractional fermion number itself remains a beckoning goal [4]. Recent breakthroughs in materials engineering such as the isolation of graphene [5] and advances in the fabrication and understanding of topological insulators [6] offer hope that this goal can be reached shortly. Among the physical mechanisms studied in this respect is the possibility of generating exotic gap terms in graphene which otherwise shows a massless Dirac dispersion relation [3, 7, 8]. The detection of fractionalization is important because it is regarded as a direct consequence of topological invariants in field theories [9] and would further vindicate the physical reality of Dirac's negative energy sea [10]. For graphene research, fractionalization may also be related to the eagerly-awaited fractional quantum Hall effect in suspended graphene sheets [11].

This letter predicts the detection of fractionalization through the zero-energy states (or zero modes) of bilayer graphene (BG) with bias. Consisting of two coupled graphitic monolayers with basis atoms $A_2$, $B_2$ and $A_1$, $B_1$ at the top and bottom layers, respectively, and arranged according to Bernal stacking ($A_2$, $A_1$) (see Fig. 1 a), BG possesses unique features such as a controllable semiconductor gap [12] and trigonal warping [13] that may be crucial for realizing fractionalization. As we verify below, topological theorems guarantee the existence of an odd number (even number including 0) of zero modes in the biased (unbiased) case. Since fractionalization is premised on an odd number of zero modes, the doubling of such modes on account of the valley degeneracy of BG will be a crucial problem. Indeed inter-valley scattering in graphene is a current subject of intense interest [14]. We propose that the degeneracy between K- and K'-valley zero modes can be lifted by a layer asymmetry responsible for charge depletion, thus leading to our conclusion that charge-½ edge states in semi-infinite BG with zigzag edge may be detected experimentally.

Let us recall the physics of fractionalization. We begin with the Dirac equation to describe the propagation of a fermion in a non-quantum background configuration. Under an ordinary potential, the fermion modes emerge as charge conjugate pairs and the coefficients of the Dirac fields are creation and annihilation operators for fermions and antifermions. In a soliton background $\varphi(x)$ or for biased BG, an extra excitation mode appears which is without a charge conjugate counterpart. The expansion coefficients for this new mode now require a pair of energy degenerate states and charge conjugation stipulates that they carry fermion number $\pm\frac{1}{2}$. Specifically, for an operator $O$, and assuming a single zero mode $\psi_0$, tr($O$) = $\langle\psi_0|O|\psi_0\rangle$ + $(\int_{E>0}+\int_{E<0})dE\langle\psi_E|O|\psi_E\rangle$. If a conjugation symmetry exists which takes positive-energy states to negative-energy ones, i.e. if an operator $\Omega$ exists which anticommutes with the Hamiltonian, then tr($O$) = $\langle\psi_0|O|\psi_0\rangle+2\int_{E<0}\langle\psi_E|O|\psi_E\rangle dE$. Suppose now that $O$ is the number operator. It is natural to define the particle number in the background field (with states $\Psi_E$) relative to the system *without* the background (with states $\psi_E$), $\rho(x) = \int_{-\infty}^{0^-}dE\{|\Psi_E|^2-|\psi_E|^2\}$. Writing the completeness relation as $\int_{-\infty}^{\infty} dE |\psi_E|^2 = |\Psi_0|^2 + \int_{-\infty}^{0^-} dE |\Psi_E|^2 + \int_{0^+}^{\infty} dE |\Psi_E|^2$, and assuming the negative energy sea is filled but the zero-energy state empty, we have $\rho(x) = -\frac{1}{2}|\Psi_0(x)|^2$. It emerges that $\Psi_0$ is normalizable, so the fermion number is $N_F = -\frac{1}{2}$. If the zero-energy state is occupied a similar calculation yields $N_F = +\frac{1}{2}$. The special property of the bosonic soliton background $\varphi(x)$ that led to fractionalization was its topological character, whereby $\varphi(+\infty)$ and $\varphi(-\infty)$ took non-equal values of opposite signs [1, 15].

We adapt a two-band description of the low-energy dynamics of BG in which the energy difference $\Delta$ between inequivalent carbon atoms is set to zero [16]:

$$H_B = \begin{pmatrix} \pm\frac{1}{2}V & -\frac{1}{T}\pi^{\mp 2} \pm v_3 \pi^{\pm} \\ -\frac{1}{T}\pi^{\pm 2} \pm v_3 \pi^{\mp} & \mp\frac{1}{2}V \end{pmatrix} \quad (1)$$

where $\pi^{\pm} = p_x \pm ip_y$, $V$ is the bias voltage, $t_\perp$ the coupling between A1 and A2 atoms, $T = t_\perp \equiv 1/2m^*$, $\varepsilon_0 \equiv \hbar^2 k_0^2/2m^*$. The quantity $v_3 \equiv \hbar k_0/2m^*$ describes the hopping B1 $\to$ B2 and has been shown to account for the trigonal warping which breaks rotational symmetry [17]. The upper (lower) sign applies to the valley K (K') and the basis in valley K (K') is $(\psi_{B2}, \psi_{B1})$ ($(\psi_{B1}, \psi_{B2})$). For the $V = 0$ case, i.e. the *unbiased* case, we use the symbol $H_0$ instead. In the vicinity of the energy $\varepsilon_0$ ($\varepsilon_0 \cong 3.9$ meV) the k-linear and k-square terms are comparable: for energies less (greater) than $\varepsilon_0$ the former (latter) dominates [17]. Two formal symmetries between valley-Hamiltonians are $H_{0\mp}^* = H_{0\pm}$ and $H_{B\mp} \xrightarrow[x \to -x]{} H_{B\pm}$ ($\pm$ refers to valleys). According to the first, for the unbiased case, K-valley states propagate in the reverse direction to the corresponding K'-valley states. The second tells us that a coordinate reflection about the *y*-axis effectively exchanges valleys so a second reflection brings us back to the same valley; but there is nothing to preclude this incurring an overall sign change in $H_B$.

Solutions of the Dirac equation $H_B \begin{pmatrix} A(r,\phi) \\ B(r,\phi) \end{pmatrix} = E \begin{pmatrix} A(r,\phi) \\ B(r,\phi) \end{pmatrix}$, expressed in polar coordinates, can be found by assuming the ansatz (*k* = wave vector, *a, b* = constants)

$$A(r,\phi) = a \sum_{m=-\infty}^{\infty} e^{i\nu\phi}(i)^m J_\nu(kr), \quad B(r,\phi) = b \sum_{n=-\infty}^{\infty} e^{i\mu\phi}(i)^n J_\mu(kr) \qquad (2)$$

If $\mu$, and $\nu$ equal the integers *n* and *m* respectively, then the above are the familiar Rayleigh expansion of a plane wave, which is single valued about the origin (even parity). However we may also assume $\mu = m + \tfrac{1}{2}, \nu = n + \tfrac{1}{2}$, in which case $A(r,\phi), B(r,\phi)$ are doubled-valued on circling the origin, i.e. $A(r,\phi+2\pi) = -A(r,\phi)$, and similarly for $B(r,\phi)$ (i.e., they are odd-parity fields). These parity choices correspond to a symmetry of $H_B$ noted above. Thus a bilayer plane wave mode has a corresponding odd-parity counterpart. We distinguish the angle between the momentum and radius vector $\phi$, from the polar angle $\theta$. Invoking the relation, $\pi^\mp = \tfrac{\hbar}{i} e^{\mp i\theta}(\nabla_r \mp \tfrac{i}{r}\nabla_\theta)$, and standard Bessel-function identities, we find the same energy dispersion for *both* choices of parity [17],

$$e^2_{K(K')} = v^2 + \xi^2\left(\xi^2 + 1 \mp 2\xi\cos 3\gamma\right) \qquad (3)$$

where $e = E/\varepsilon_0, \xi = k/k_0$ and $v = V/2\varepsilon_0$ are dimensionless, $\gamma = \theta - \phi$ and the upper (lower) sign applies to the valley K (K'). Although the trigonal interaction leads to asymmetry in the valley energies, $e_{K(K')}(k) \neq e_{K(K')}(-k)$, time-reversal symmetry still holds since $e_{K(K')}(k) = e_{K'(K)}(-k)$. We have also the following relations,

$$a = e^{i\vartheta_K}, \quad b = \frac{\frac{1}{2}V - E}{\varepsilon_0 \frac{k}{k_0^2}\sqrt{k^2 + k_0^2 - 2kk_0\cos 3\gamma}} s, \quad \vartheta_K = \arg(ke^{-2i\gamma} - k_0 e^{i\gamma}), \quad E = s|E_K|, \quad s = \pm 1 \quad (4)$$

The corresponding result for K' is given by replacing $\vartheta_K$ with $\vartheta_{K'} = \arg(ke^{2i\gamma'} + k_0 e^{-i\gamma'})$. For identical energies and wave numbers, we have $\gamma' = \gamma + \pi/3$. Contour plots of the dispersion are displayed in Fig. 1 b. The $k \to -k$ asymmetry within a valley is clear. For large $k$ the unbiased and biased cases become gradually indistinguishable from each other. By increasing the bias we are able to explore the $k$-linear regime at higher particle momenta. This observation is important for zero modes which are essentially $k$-linear modes. The trigonal interaction is also important here because it is responsible for generating the zero modes at equiangular directions (see arrows in Fig 1 c), thus allowing us to examine one of these modes and its corresponding charge.

For unbiased BG, four zero-energy modes were found to occur at $k = 0, k_0 e^{i\gamma}$ where $\gamma = 0, \pm 2\pi/3$ as shown in Fig. 1 (b). The K-valley wave functions for nonzero $k$ (chosen to lie on the $x$ axis) are exemplified by the plane-wave pair $\Psi_{0,\pm,K} = \binom{1}{0} e^{ik_0 x}$, and $\binom{0}{1} e^{-ik_0 x}$. More generally, in the unbiased case the Hamiltonian $H_0$ satisfies the equation $\sigma_z H_0 \sigma_z = -H_0$ which expresses a conjugation symmetry in which positive-energy solutions map to negative energy solutions and vice versa, so the fermion number density is an even function of $E$. This case is thus particle-hole symmetric and the zero-mode pair in particular, which obeys $\sigma_z \Psi_{0,\pm,K} = \pm \Psi_{0,\pm,K}$, is said to be chirally conjugate (the chiral conjugate $\Psi^c_{0,\pm,K} = \sigma_z \Psi^*_{0,\pm,K}$ is associated with the negative energy). Observe that while every nonzero-energy state has two solutions propagating in a given direction (i.e. Eq. (4)), there is only one zero mode moving in a given direction; that is, there are half as many zero modes as modes of any other non-zero energy state. But the zero modes $\Psi_{0,\pm,K}$ are not localized. Similar remarks apply to the valley K'.

Let us now examine the winding number $N$ given by the formula $N = \frac{1}{4\pi i}\int_0^{2\pi} d\theta \, \text{Tr}(\sigma_z H^{-1}\partial_\theta H)$ [18]. For unbiased BG, the Hamiltonian has the generic form $H_0 = \begin{pmatrix} 0 & p \\ p^* & 0 \end{pmatrix}$. From the symmetry of $H_0$ the Hamiltonian for negative energies is $H_0^*$ and it follows that in computing $N$, the negative energy states cancel the contribution from the positive states. Thus the only non-cancelling contributions to $N$ can come from the zero modes and $N$ will thus give a good indication of their number, as we now show. The Hamiltonian around the Fermi K point, $H_0(K+k)$, for instance, furnishes us with a map from the circle $k_x^2 + k_y^2 = R^2$ to the space of $2\times 2$ Hamiltonians $H_0 = \boldsymbol{h}\cdot\boldsymbol{\sigma}$. Since Fermi points correspond to zeros of the determinant - $\det(H_0) = h_x^2 + h_y^2$, a Fermi point may be associated putatively with one or more zero modes. We infer that the contribution to $N$ of a given single-valued zero mode $\Psi_0 = \binom{a}{b}$ is $\pm 1$. (This is because $H_0^{-1}\partial_\theta H_0|\Psi_0\rangle = -|\partial_\theta \Psi_0\rangle$ so $\langle\Psi_0|\sigma_z H_0^{-1}\partial_\theta H_0|\Psi_0\rangle = -\frac{1}{2}\partial_\theta\langle\Psi_0|\sigma_z|\Psi_0\rangle = -\partial_\theta|a|^2$.) Two Fermi points with opposite contributions to $N$ implies that they mutually annihilate each other, since one 'undoes' the contribution of the other. Hence it is the net sum of the contributions that gives the total number of zero modes. For $H_0$ we find ($x = p/Tv_3 \geq 0$),

$$N = -\int \frac{d\theta}{2\pi} \frac{-2x^4 + x^2 + x^3 \cos 3\theta}{x^4 + x^2 - 2x^3 \cos 3\theta},$$ with a Fermi point occurring at $x = 0$ and a Fermi 'triangle' at $x = e^{\pm 2i\pi/3}, 1$. The former has a contribution of -1 and the latter three contribute collectively +3, yielding the total winding number 2, confirming the existence of the four zero modes as shown in Fig.1 (b). Because of this even number, unbiased BG then is unsuitable for detecting fractionalization.

Before addressing the biased case, we first observe that the above winding-number interpretation applies to anti-symmetric Hamiltonians, which is not true of the biased case. Instead we will resort to calculating the topological charge $N_3$ of the associated Fermi point, which can be interpreted as the difference between the number of left-moving and right-moving zero modes [19]. Regarding the biased Hamiltonian $H_B$ as a vector in the space of Pauli matrices, this is given by ($\zeta \equiv b/a^2$ is dimensionless)

$$4\pi N_3 \equiv \int \frac{\vec{H}_B \cdot \nabla_{p_x} \vec{H}_B \times \nabla_{p_y} \vec{H}_B}{|\vec{H}_B|^3} d^2p = 2b \int d^2p \frac{p^2 - \tfrac{1}{4}a^2}{\{p^4 + a^2 p^2 - 2ap_x(p_x^2 - 3p_y^2) + \tfrac{1}{4}b^2\}^{3/2}}$$
$$= 4\sqrt{\pi}\zeta \sum_n \frac{\Gamma(2n+\tfrac{3}{2})}{\Gamma(1+n)^2} \int dx\, x^{3n} \frac{x - \tfrac{1}{4}}{(x^2 + x + \tfrac{1}{4}\zeta^2)^{2n+3/2}} \tag{5}$$

The integration was carried out by first writing the denominator as a three-dimensional Gaussian integral. Numerical evaluation of the last result yields $N_3 = 1$, in agreement with the earlier predictions of an odd number (three) of zero modes, as shown in Fig.1 (c).

To construct suitable localized zero modes for an *infinite* graphene sheet in the *biased* case, we first rescale $X(Y) = k_0 x (k_0 y)$, let $A(x,y) = e^{ipX} a(Y)$, $B(x,y) = e^{ipX} b(Y)$ (with $p$ positive), and assume further that $a(Y)$ and $b(Y)$ have exponential form: $a(Y) = a e^{\lambda Y}$. Substitution into Eq. (1) and demanding consistency implies

$$0 = v^2 + (\lambda^2 - p^2)^2 - (\lambda^2 - p^2) \mp 6p\lambda^2 \mp 2p^3 \tag{6}$$

The upper (lower) sign applies to the K (K') valley. There are four roots: one pair applies to the wave function for $x > 0$ and the other to $x < 0$. To determine the coefficients of the exponential solutions in the full expressions for $A(x, y)$ and $B(x, y)$ we demand continuity of the wave function and its first derivative as well as the validity of Eq. (1) at the origin. Typical wave functions are displayed in Fig. 2. Although the sublattice densities may be quite similar, the last row shows that the valley densities can be quite dissimilar. Two other zero modes can be constructed by simply rotating the above mode by angles $\pm 2\pi/3$ (Fig 1 c). These satisfy the same dispersion relation (6).

However, for the purpose of detection, we consider now semi-infinite BG with a zigzag edge at $y = 0$, such that the appropriate boundary condition is $B(x, y = 0) = 0$. For a given $p$-value, i.e., $x$-momentum and bias $v$ and writing $\lambda = \alpha + i\beta$, the K-valley solution has the form $B_p(x, y) = c_0 e^{ipX - |\alpha|Y} \sin \beta Y$, $A_p(x, y) = c_1 e^{ipX - |\alpha|Y} \sin(\beta Y + \theta)$, where $v c_1 / c_0 = (p - \lambda_1)^2 - (p + \lambda_1)$ and $\theta =$ phase ($c_1$). The corresponding K'-valley result is given by replacing $c_1$ and $\theta$ by $d_1$ and $\theta'$, where $v d_1 / d_0 = v / [(p + \lambda_1)^2 + (p - \lambda_1)]$. While in the unbiased case we had four zero-energy solutions per valley, in the present case there are only three per valley. Of particular interest here is the one with topological charge of -1 (see Fig.1(c)) which can be inferred to be localized with peak at the $y = 0$ axis (zigzag edge). Predicting the existence of a localized edge state zero mode in biased BG and attributing it to charge fractionalization is the central result of this paper.

Formally, we can show that $\Omega H_B^T \Omega = H_B$, where $\Omega = i\sigma_y$. It follows that $\psi^T \Omega H_B = -E\,\psi^T \Omega$. Then we can verify that the zero-energy soluton is indeed self-conjugate. The nontrivial topology here which was absent in the unbiased case is the different signs of the bias for the top and bottom layers of BG. Compared with the infinite sheet, the interesting new feature of the semi-infinite geometry is the possibility of having almost all the charge residing only on one sublattice and at the zigzag edge. As Fig. 3 shows, the B2 component is much smaller than the B1 component for both valleys, with the difference being more pronounced for the K-valley component. Clearly the charge becomes more concentrated on one edge as bias and *x*-momentum are increased. This should be useful in the detection of fractional charge. In principle two other zero modes are present (with topological charge +1 each in Fig. 1 c) but since these do not reside at the zigzag edge their existence need not affect this result. They have similar profile to the infinite-sheet zero modes.

Nevertheless, to observe fractionalization unambiguously, one still needs to distinguish the zero modes from two different valleys for these two modes occur spatially at the same zigzag edge. Hence we consider an additional perturbation whose effect on Eq. (1) is an extra term, $\pm \dfrac{V}{2}\dfrac{1}{\gamma_1 m^*}\begin{pmatrix} \pi^+\pi^- & 0 \\ 0 & -\pi^-\pi^+ \end{pmatrix}$, which describes a layer asymmetry due to the depleted charge on the B sites [20]. The plus (minus) sign is for Dirac point K (K'). Typically, $\gamma_1 \cong 0.4\,\text{eV}$ [12]. This perturbation clearly splits the valley degeneracy in opposite directions. Since we already know that the B1 component of the wave function is generally dominant we will simply calculate its contribution, neglecting the other component. To first order in perturbation theory, we estimate the energy shift between the zero modes

$$\Delta E_0 \cong \pm \varepsilon_0 \frac{v \hbar^2 k_0^2}{\gamma_1 2 m^*} \int dx dy \left| \nabla A_p(x, y) \right|^2 \approx \mp \frac{V}{\gamma_1} \varepsilon_0 \left| \lambda - p \right|^2 . \qquad (7)$$

Although the ratio $(V/\gamma_1)$ is typically of order 0.1, the quantity $\left| \lambda - p \right|^2$ can be much larger than unity ($O(10^2)$ for $v = 50$, $p = 9$ in Fig. 3) so $\Delta E_0$ can be large enough to clearly discriminate the valley modes.

In summary, we have studied BG, and showed that in the unbiased case the valley zero modes occur in pairs, rendering them unsuitable for detecting fractionalization. In the biased case, we noted first that the presence of the bias lent us the freedom to explore k-linear phenomena at higher particle momenta, which is the mechanism responsible for zero modes. Moreover the bias played the role of a bosonic field with nontrivial topology. For semi-infinite BG with zigzag edge we constructed the self-conjugate and localized (edge-state) zero mode. Although the K'-valley also gave rise to a similar zero mode, the presence of a layer asymmetry lifted this degeneracy thereby allowing for the unambiguous detection of charged-½ states in each valley.

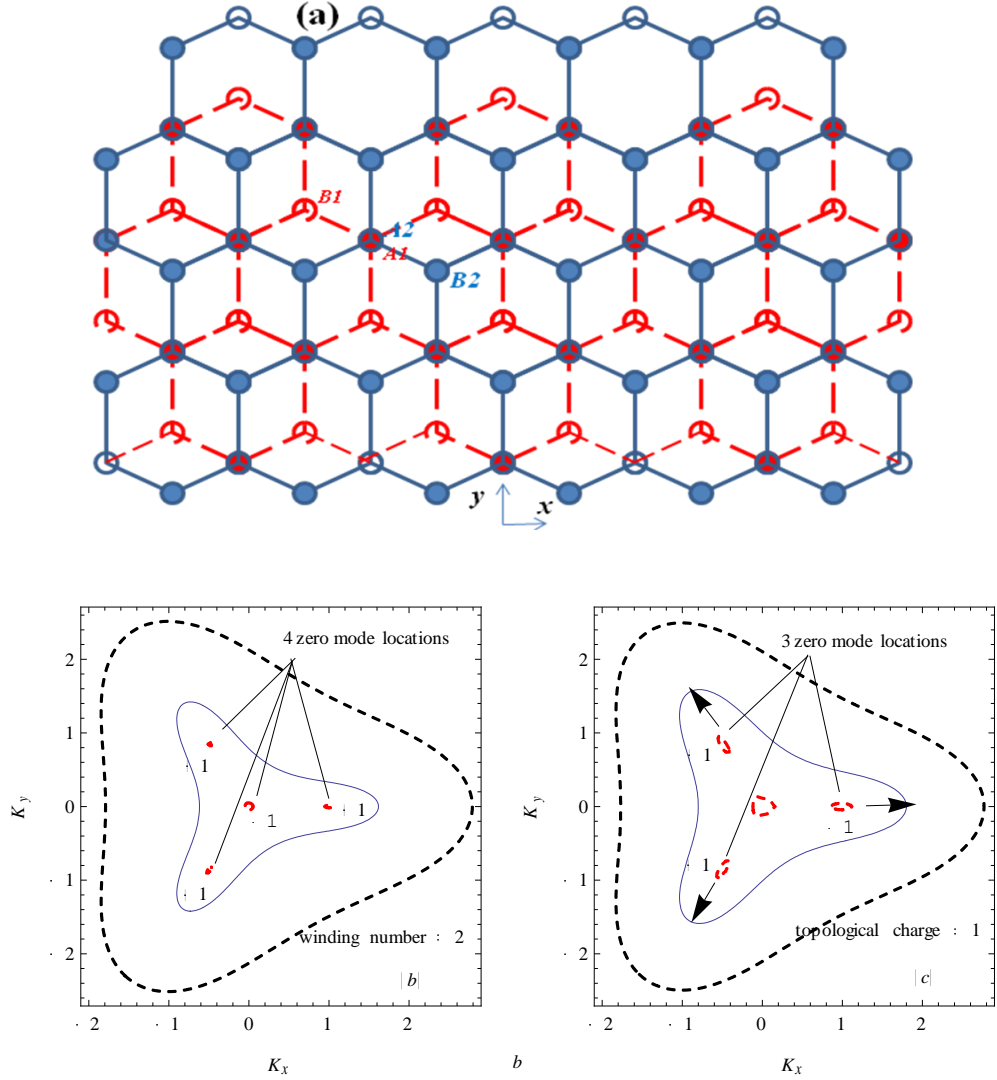

**Fig. 1** (a) Bilayer graphene ribbon with zigzag edges. The A2 atom is on top of the A1 atom. The ribbon is periodic in the $x$-direction and of length $L$ along the $y$ direction. (b) Equi-energy contours around the K point for the unbiased case (left) for $e = 0.1$ (dashed), 1 (solid), and 5 (dark, dashed); (c) for the biased ($v = 4.996$) case (right) for $e = 4.9978$ (dashed), 5.2 (solid) and 7 (dark, dashed). Note $K_x = k_x/k_0$, etc. Units are discussed after Eq. (3). Other details are discussed in the text.

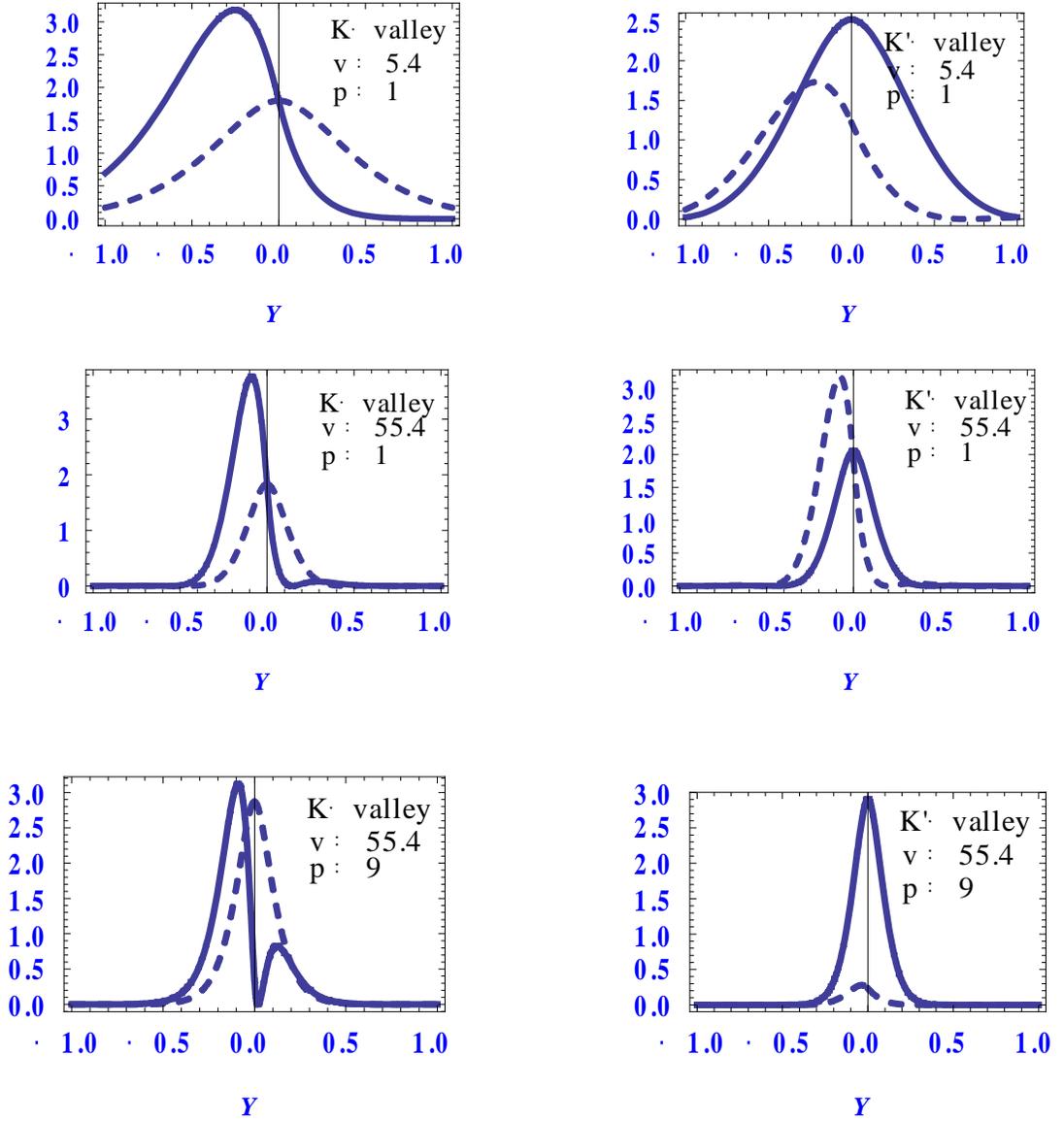

**Fig. 2** Graphs of the unnormalized wave functions for an infinite graphene sheet: $|A(X,Y)|^2$ (dark) and $|B(X,Y)|^2$ (dashed) for the K-valley (left column) and K'-valley (right column) for $v = 5.4$, $p = 1$; $v = 55.4$, $p = 1$; $v = 55.4$, $p = 9$. Vertical units are arbitrary. $Y = 1$ corresponds to 17 nm.

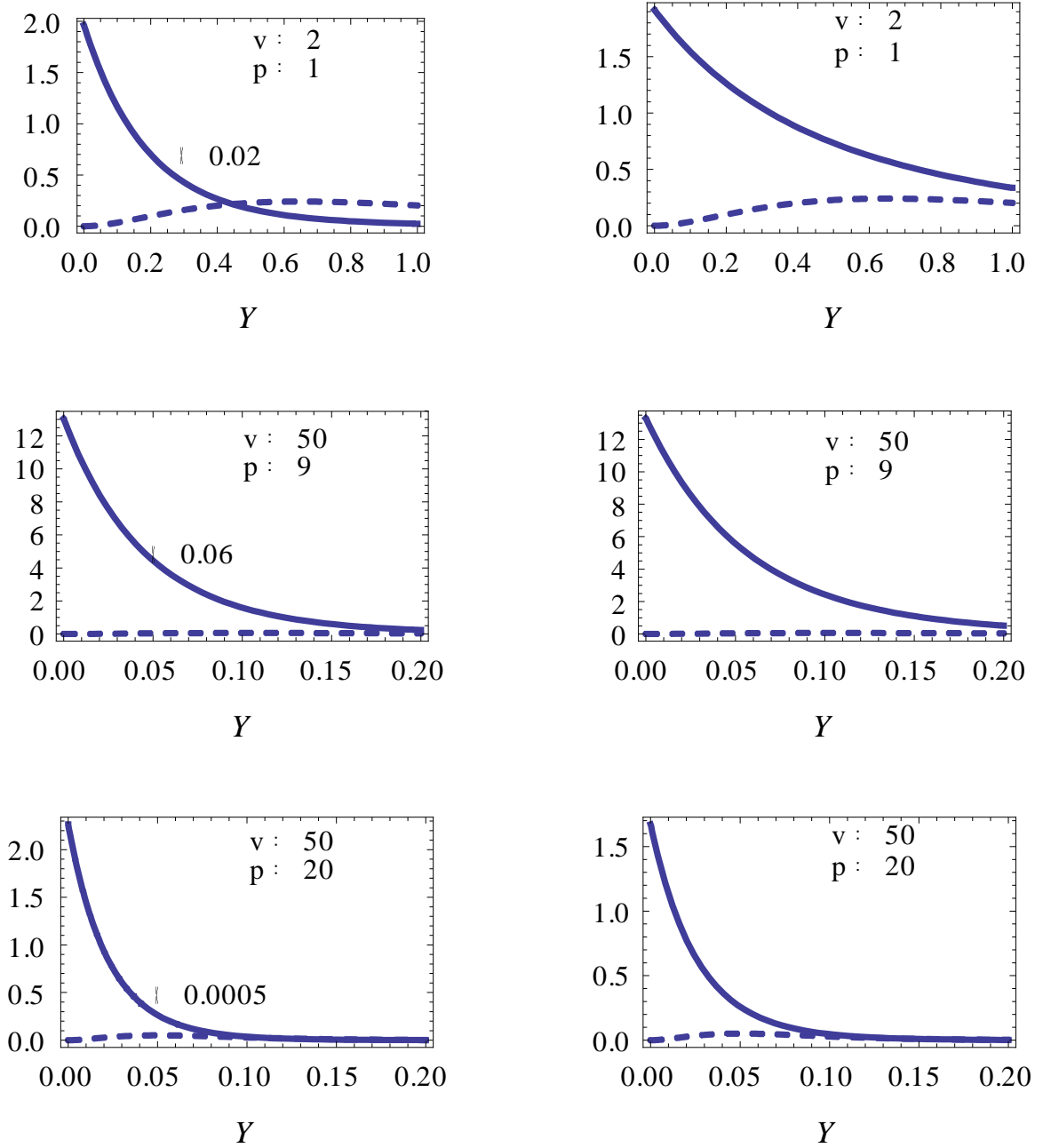

**Fig. 3** Graphs of unnormalized wave functions for a half-infinite sheet with zigzag edge: $|A(X,Y)|^2$ (bold) and $|B(X,Y)|^2$ (dashed) for different bias voltage and x-momentum. The left column is for the K-valley and the right for the K'-valley. Scales are the same as Fig. 2. Note that the left-column $A(X,Y)$ graphs are multiplied by a numerical scale factor. Thus graphs show that $B(X,Y)$ is much smaller than $A(X,Y)$ specially for the K-valley. For larger bias and x-momentum the wave functions are more concentrated at the edge. This is an ideal scenario for the detection of charge fractionalization.